\documentclass[letterpaper]{achemso}

\usepackage[T1]{fontenc}

\usepackage{geometry}
\usepackage{setspace}

\usepackage{achemso}

\usepackage{graphicx}
\usepackage{float}
\newfloat{scheme}{htbp}{los}
\floatname{scheme}{Scheme}
\floatname{chart}{Chart}
\newfloat{graph}{htbp}{loh}

\usepackage{chemformula} 
\usepackage[version = 4]{mhchem} 

\author{Felix Eder}
\affiliation[Unige]
{University of Geneva, Department of Quantum Matter Physics, 24 Quai Ernest-Ansermet, CH-1211 Geneva, Switzerland}
\author{Zeno Maesen}
\affiliation[PSI]
{PSI Center for Neutron and Muon Sciences, Forschungsstrasse 111, CH-5232 Villigen, PSI, Switzerland}
\author{Yurii Skourski}
\affiliation[HZDR]{Dresden High Magnetic Field Laboratory (HLD-EMFL), Helmholtz-Zentrum Dresden-Rossendorf, Bautzner Landstraße 400, 01328 Dresden, Germany}
\author{Enrico Giannini}
\affiliation[Unige]
{University of Geneva, Department of Quantum Matter Physics, 24 Quai Ernest-Ansermet, CH-1211 Geneva, Switzerland}
\author{Oksana Zaharko}
\affiliation[PSI]
{PSI Center for Neutron and Muon Sciences, Forschungsstrasse 111, CH-5232 Villigen, PSI, Switzerland}
\author{Fabian O. von Rohr}
\affiliation[Unige]
{University of Geneva, Department of Quantum Matter Physics, 24 Quai Ernest-Ansermet, CH-1211 Geneva, Switzerland}

\title{Resolving growth-induced off-stoichiometry in \ce{AgCrSe2} single crystals}

\begin{document}

\maketitle

\begin{abstract}
 The layered delafossite-like antiferromagnet \ce{AgCrSe2} is a superionic conductor at high temperatures and has been reported to exhibit anomalous Hall behavior and Kondo physics at low temperatures. These extraordinary transport properties have been established almost exclusively on single crystals grown by chemical vapor transport, raising questions about the role of growth-induced off-stoichiometry. Using elemental analysis, single-crystal X-ray diffraction, and magnetization measurements, we show that such crystals are indeed systematically off-stoichiometric, with a general composition of \ce{Ag_{1-\textit{x}}Cr(Se_{2-\textit{y}}Cl_{\textit{y}})} ($x \approx y \approx 0.08$) arising from the use of \ce{CrCl3} as a transport agent. This off-stoichiometry manifests in altered magnetic properties, most notably a suppressed Néel temperature of 46\,K compared to 58 K in stoichiometric polycrystalline samples prepared by solid-state synthesis. By optimizing an Ag/Se self-flux growth method, we obtained large single crystals of \ce{AgCrSe2} that recover the magnetic transition temperature and saturation field of stoichiometric powder samples. These results establish self-flux growth as a route to high-quality stoichiometric \ce{AgCrSe2} single crystals and provide a reliable platform for reassessing whether the reported anomalous transport phenomena are intrinsic or arise from off-stoichiometry.
\end{abstract}

\section*{Keywords}

Crystal Growth, Self-Flux, Magnetism, Delafossites


\section{Introduction}

The delafossite family of compounds with a composition of \textit{AMX}\textsubscript{2} is known to host a wide range of emergent properties, including multiferroicity \cite{Singh2009,Lopes2011}, superionic conduction\cite{Ag_Murphy1977}, and unconventional superconductivity.\cite{takada2003superconductivity} Many of these properties can be traced back to its layered crystal structure, consisting of \ce{\textit{MX}2} layers formed from transition metal \textit{M} atoms octahedrally coordinated by chalcogenide \textit{X} anions, which are intercalated by sheets of monovalent cations \textit{A}. As the \textit{M} atoms form a two-dimensional triangular lattice within the layer, magnetic \textit{M} species can easily introduce magnetic frustration and enable a multitude of magnetic states and transitions, which are influenced by the type, size and amount of \textit{A}, as well as the resulting structure type.\cite{nocerino2023competition,Song2021,Eder2025}

One prominent delafossite-type phase that has attracted significant attention during the past years due to its rich transport behavior, is the layered semiconductor \ce{AgCrSe2}. At high temperatures (\textit{T}\textsubscript{trans} = 475\,K), \ce{AgCrSe2} undergoes a phase transition, which manifests itself in a disordering of the Ag cations over all possible interlayer tetrahedrally coordinated sites. \cite{Ag_vanderLee1989, Ag_wakamura1990} This disorder of Ag cations above the transition temperature then leads to the establishment of a superionic state.\cite{Ag_Murphy1977,Ag_gascoin2011,Ag_Ding2020}. 
At low temperatures, Kondo transport with a Kondo temperature of 32 K and anomalous Hall effect have been reported recently. \cite{Ag_guimaraes2024,Ag_kim2024_O} 

Most of these advanced investigations of the multifarious transport properties of \ce{AgCrSe2} have been enabled by the growth of large single crystals through chemical vapor transport (CVT). Starting from polycrystalline \ce{AgCrSe2}, which can be prepared from the elements at temperatures of 900--1000 °C\cite{ag_hahn1957_first}, bromine \cite{Ag_vanderLee1989} and especially \ce{CrCl3} \cite{Ag_wakamura1990,Ag_yano2016crystal,Baenitz2021} have been established as primary transport agents in this process. 

However, one crucial aspect in this regard is the actual composition of the large CVT-grown \ce{AgCrSe2} crystals used in recent investigations. A paradigmatic example of the sensitivity of delafossite-type compounds to \textit{A}-site occupancy is provided by \ce{Na_{\textit{x}}CoO2}, in which subtle changes in the Na content lead to dramatic modifications of the electronic, magnetic, and thermoelectric properties, including charge ordering, magnetic instabilities, and unconventional superconductivity upon hydration.\cite{schaak2003superconductivity,foo2004charge} This pronounced dependence on the concentration and ordering of the interlayer cations highlights that deviations from ideal stoichiometry in layered delafossites are not a minor perturbation, but can fundamentally alter the underlying ground state.

Baenitz \textit{et al}.\cite{Baenitz2021} state that their samples tend to exhibit off-stoichiometry with an average composition of \ce{Ag_{0.92}Cr_{1.08}Se2}. The pronounced presence of defects is known to influence the transport properties of \ce{AgCrSe2} \cite{Ag_hua2021tuning}, and the recently discovered anomalous Hall effect is believed to be linked to the off-stoichiometry as well \cite{Ag_kim2024_O}. 
Furthermore, a discrepancy in the Néel temperature between polycrystalline material from solid-state synthesis and CVT-based crystals can be observed. For polycrystalline samples, initial investigations, limited on very few data points yielded a Néel temperature \textit{T}\textsubscript{N} of $\sim$50 K \cite{Bongers1968} paired with an asymptotic Curie-Weiss temperature $\theta$\textsubscript{CW} of 72 K. Later reinvestigation of polycrystalline \ce{AgCrSe2} \cite{Ag_gautam2002,Ag_Qi2025} determined \textit{T}\textsubscript{N} to be 55 K with a $\theta$\textsubscript{CW} of 88 K. Based on neutron powder diffraction, the magnetic structure was determined to be a layered antiferromagnet with incommensurate helimagnetic ordering \cite{engelsman1973} of the magnetic moments, with the moments oriented within the \ce{CrSe2} planes. In contrast, samples grown from the gas phase consequently exhibit a lower Néel temperature of $\sim$46 K \cite{Baenitz2021,Ag_han2022,Ag_guimaraes2024,Ag_kim2024_O}. 

In light of the widespread reliance on gas-phase--grown \ce{AgCrSe2} crystals that exhibit off-stoichiometry and magnetic properties inconsistent with polycrystalline material from solid-state synthesis, an alternative crystal-growth approach is required. Self-flux growth represents a well-established alternative to chemical vapor transport and has enabled, for example, polytype-selective growth of transition metal dichalcogenides such as \ce{MoTe2} and \ce{WSe2} \cite{guguchia2018,Gustafsson2018,Edelberg2019} as well as the isolation of the first monolayer ferromagnet \ce{Cr2Ge2Te6}. \cite{ji2013,gong2017} Building on recent work demonstrating the applicability of self-flux growth to delafossite-type chalcogenides,\cite{witteveen2023synthesis,Eder2025} as well as the self-flux controlled stoichiometry of these phases,\cite{eder2025stoichiometry} we identify an Ag/Se self-flux method combined with hot centrifugation as a suitable route to obtain \ce{AgCrSe2} single crystals.

In this work, we report the growth of large, stoichiometric \ce{AgCrSe2} single crystals using an Ag/Se self-flux method followed by hot centrifugation. For comparison, we also prepared \ce{AgCrSe2} via conventional solid-state synthesis, chemical vapor transport, and self-selective vapor growth, enabling a systematic comparison of composition and magnetic properties across different growth routes.

\section{Experimental}
\subsection{Synthesis}
All reactions were performed from the same batches of chemicals, which were used as purchased: silver (powder, Alfa Aesar, 99.9\%), chromium (powder, Alfa Aesar, 99.99\%), and selenium (pieces, Alfa Aesar, 99.999\%). For crystals grown from the gas phase, \ce{CrCl3} was used as transport agent. 

Polycrystalline \ce{AgCrSe2} powder was synthesized by mixing and co-grinding stoichiometric amounts of the elements, sealing them in a quartz ampule with a partial Ar atmosphere of 300 mbar, and tempering the setup at 800 °C for 96 h with heating and cooling rates of 60 and 20 °C/h, respectively. 
CVT crystals were grown following the recent literature procedure \cite{Baenitz2021}, albeit with a reduced reaction time of two weeks.
For self-selective vapor growth (SSVG), 500 mg polycrystalline \ce{AgCrSe2} powder was mixed with some \ce{CrCl3} to achieve a \ce{Cl2} concentration of 3 mg/cm\textsuperscript{3}, like in the CVT experiment, inside an evacuated quartz ampule. Within a muffle furnace, the ampules were heated up at a rate of 120 °C/h to 925 °C, which was then held for 120 h. After this, the temperature was slowly reduced at 1\,°C/h to 825 °C, which was subsequently held for another 72 h, before the furnace was cooled to room temperature at 120 °C/h. The experimental setup, including a flipping of the muffle furnace to its side to optimize the small internal temperature gradients, was adapted from Ref. \citenum{lukovkina2024}.

For self-flux experiments, elemental Ag, Cr and Se were weighed in in the corresponding molar ratios into \ce{Al2O3} crucibles, which were placed inside a quartz ampule and covered by a 2--3 cm thick layer of quartz wool. The quartz wool hereby operates as the sieve in the commonly used Canfield crucible assembly \cite{Canfield2016} to hold off excess flux during the hot-centrifugation step. The ampules were sealed off under 300 mbar partial pressure of Ar and heated to 1000 °C at 60 °C/h. After 48 h at maximum temperature, the reaction was slowly cooled down to 700 °C at 3 °C/h.
After four hours at this temperature, the ampules were taken out of the furnace, flipped around, and the residual Ag/Se flux was quickly removed through hot-centrifugation.
Following this protocol, we have grown crystals from various flux compositions of (given as molar Ag:Cr:Se ratios) \textit{n}:1:8 (\textit{n} = 1, 2, 3, 4, 5, 6), and \textit{n}:1:16 (\textit{n} = 6, 8) to check the compositional window for the growth of \ce{AgCrSe2}.

\subsection{Characterization}

Powder X-ray diffraction (PXRD) patterns of reaction products were collected on a Rigaku SmartLabXE diffractometer equipped with a D/teX Ultra 250 detector using Cu-K\textsubscript{$\alpha$} radiation in a recording range of 5--80° 2$\theta$. To mitigate the effects of preferred orientation, measurements were conducted in Debye-Scherrer geometry inside of quartz capillaries with an outer diameter of 500 \textmu m. To avoid excessive sample absorption, the samples were ground together and diluted with amorphous \ce{SiO2} powder.
Rietveld refinements of PXRD patterns were conducted with Jana2020 \cite{Jana2020}, the background was subtracted after completion of the refinement for clarity.

Single crystal X-ray diffraction (SXRD) experiments were conducted on a Rigaku Supernova diffractometer using Mo-K\textsubscript{$\alpha$} radiation. Unit-cell indexation, integration, and absorption correction were performed with CrysalisPro\cite{CrysAlisPRO2022}, while structure solution and refinement were done with SHELXT and SHELXL, respectively.\cite{sheldrick_shelxt_2015,sheldrick_shelxl_2015}

To quantify the eventual off-stoichiometry of the grown crystals, their elemental composition was checked by energy dispersive X-ray spectroscopy (EDS) on a JEOL JSM-7600F scanning electron microscope (SEM), equipped with an X-Max\textsuperscript{N} 80 detector (Oxford Instruments) and at an accelerating voltage of 20 kV.

Magnetic properties were investigated in a cryogen-free Quantum Design Dynacool Physical Property Measurement System (PPMS). Crystals were mounted in Kapton envelopes and oriented in two different orientations with the external field either perpendicular (\textbf{H} $\parallel$ \textbf{c}) or parallel (\textbf{H} $\parallel$ \textbf{ab}) to the morphologic layer plane. 
This macroscopic layer plane of the plate-shaped crystals coincides with the crystallographic (001) plane, as was confirmed through test-measuring about a dozen crystals of varying sizes by SXRD.
\ce{AgCrSe2} synthesized from solid-state reaction was measured in powder form, since no large enough crystals were available.
Field-dependent measurements were conducted in sweep mode in the --9 to 9 T range, temperature-dependent investigations in the 2--300 K interval. 
High-field magnetization measurements were conducted at 1.7 K under pulsed magnetic fields up to 30 T in the Dresden High Magnetic Field Laboratory (HLD) in the Helmholtz-Zentrum Dresden-Rossendorf.
In-house \textit{m}(\textit{H}) curves collected at 2 K in the 0--9 T range were used to calibrate the magnetic moment of the high-field measurements. 

\section{Results and Discussion}
\subsection{Crystal growth}

\begin{figure*}
\begin{center}
\includegraphics[width=16.5cm]{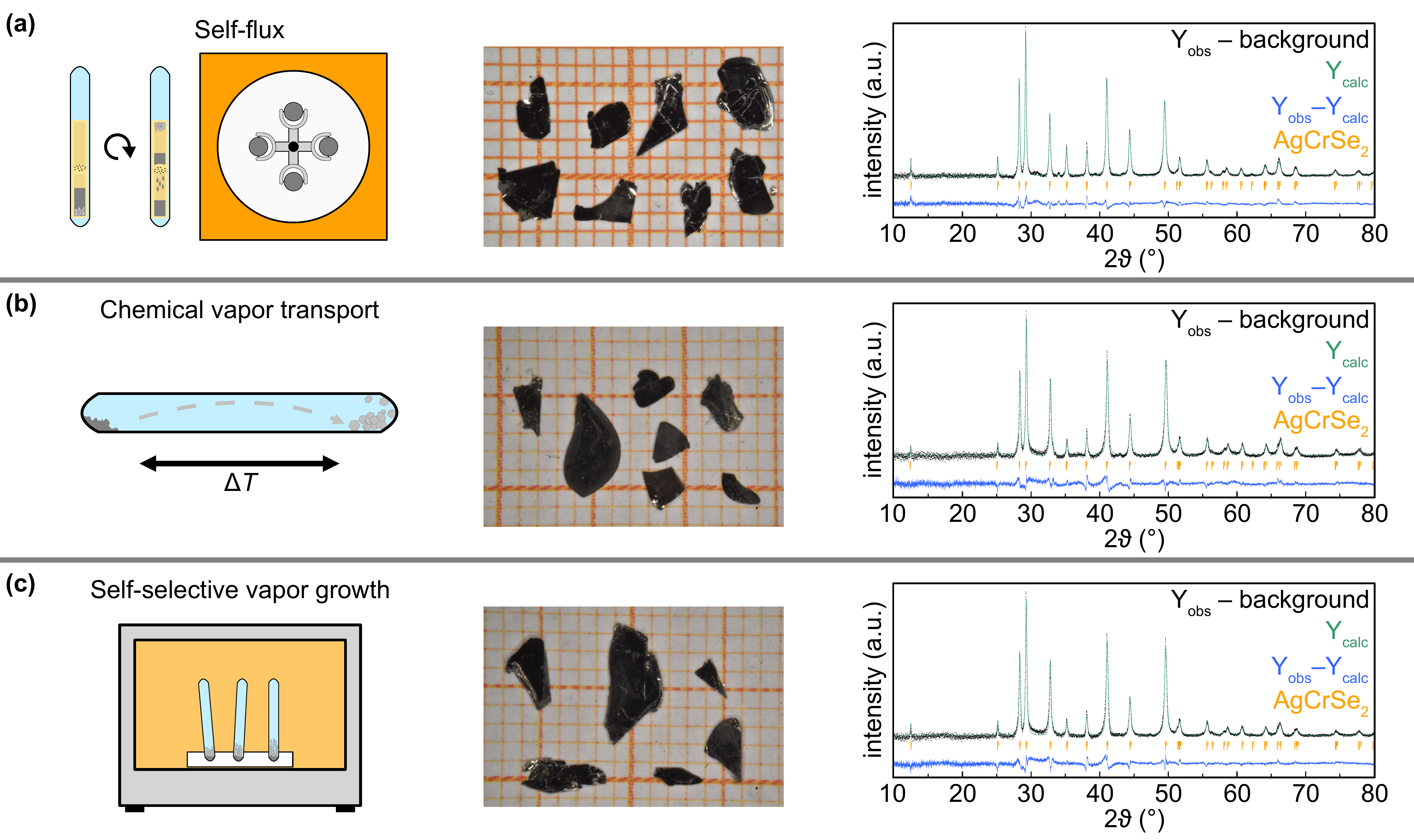}
\caption{The crystal growth of large \ce{AgCrSe2} crystals via (a) self-flux, (b) chemical vapor transport and (c) self-selective vapor growth. From left to right, schematics of crystal growth method, examples of large crystal plates on mm-paper for scaling, and Rietveld-refined PXRD patterns of thoroughly ground crystal plates are shown.}
\label{Fig1}
\end{center}
\end{figure*}

We first examine the crystal growth, morphology, and phase purity of \ce{AgCrSe2} obtained by different synthesis routes. Self-flux grown \ce{AgCrSe2} crystallizes as silver reflective plates with lateral dimensions of up to several millimeters (Fig. \ref{Fig1}(a)), which were frequently intergrown with each other. 
Among the explored compositions, flux ratios of 4:1:8, as well as the more dilute mixtures 6:1:16 and 8:1:16, yielded the best-shaped crystals with the lowest degree of intergrowth. 
\ce{AgCrSe2} was found to form in all conducted flux-reactions, and only in few cases traces of residual flux materials (\ce{Ag2Se}, Se) were detected as side products. No Ag-depleted phases, as they are known in similar alkali metal (\textit{A}) chromium selenide or sulfide systems with compositions of \ce{\textit{A}Cr5Se8} or \ce{\textit{A}Cr3S5} were found.\cite{huster1978,quint1984,bronger1993} 
Neither could any structurally different understoichiometric \ce{Ag_{1-\textit{x}}CrSe2} phases, as is, e.g., the case for \ce{K_{0.6--0.8}CrSe2} and K\textsubscript{1-\textit{x}}CrSe\textsubscript{2} (\textit{x} $\approx$ 0.13) \cite{Wiegers1980,Eder2025}, be observed.

Reference samples grown by CVT and SSVG likewise form plate-like crystals with lateral dimensions of several millimeters (Fig. \ref{Fig1}(b,c)). In PXRD experiments of ground crystals, products of the different growth methods behave similarly overall. Small differences in the lattice parameters are observed between flux-grown \ce{AgCrSe2} (\textit{a} = 3.68716(7) Å, \textit{c} = 21.2595(5) Å) and samples obtained by CVT (\textit{a} = 3.67478(9) Å, \textit{c} = 21.2656(8) Å) or SSVG (\textit{a} = 3.67663(9) Å, \textit{c} = 21.2652(8) Å). Additionally, the intensity distribution of the flux-grown samples varies a bit from their gas-phase based counterparts, as it can be seen best from the pair of reflections at 33 and 36° 2$\theta$. 
These results demonstrate that self-flux growth yields phase-pure \ce{AgCrSe2} crystals with morphology and average crystal structure comparable to gas-phase grown samples.

\subsection{Magnetic properties}

\begin{figure*}
\begin{center}
\includegraphics[width=16.5cm]{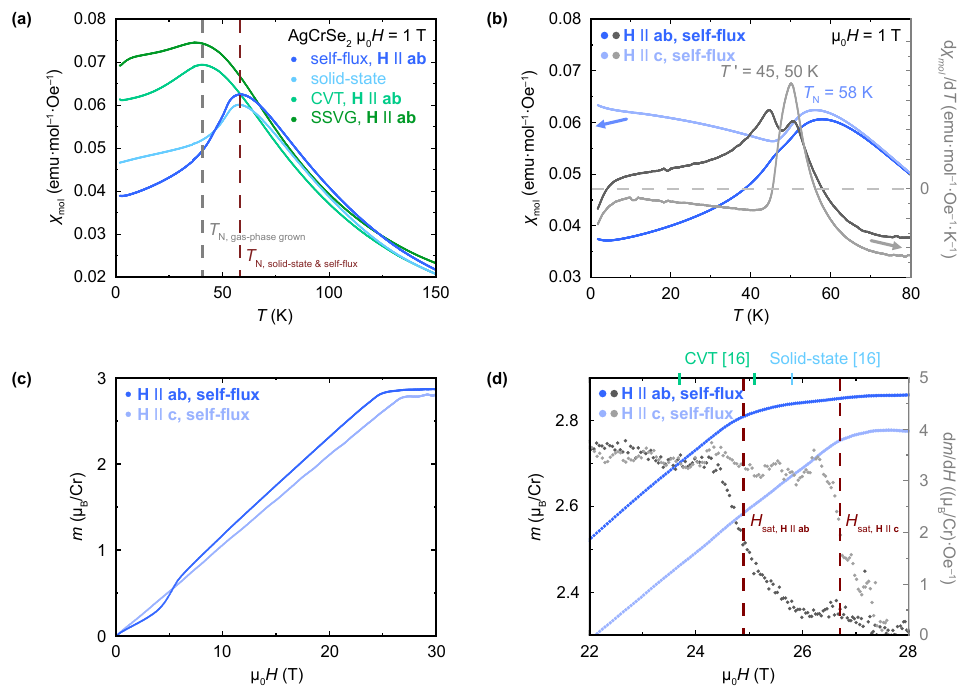}
\caption{(a) Temperature-dependent magnetic susceptibility of \ce{AgCrSe2} samples prepared by different synthesis methods with the external field of 1 T applied parallel to the layer plane; (b) The first derivative of molar susceptibility of flux-grown \ce{AgCrSe2}, highlighting the increased Néel temperature; (c) Field-dependent magnetic moment of flux-grown \ce{AgCrSe2} measured at fields up to 30 T. (d) Saturation region of \textit{m}(\textit{H}) including the first derivative in grey. Reference data on CVT and solid-state grown samples\cite{Baenitz2021} are included at the top.}
\label{Fig2}
\end{center}
\end{figure*}

To assess whether self-flux growth restores the intrinsic magnetic behavior of stoichiometric \ce{AgCrSe2}, we compare the anisotropic magnetic properties of crystals grown by different synthesis routes. 
When analyzing the temperature-dependence of the molar susceptibility $\chi_{mol}$ (here defined as $\chi_{mol} = \frac{m}{H*n}$; \textit{m} = magnetic moment, \textit{H} = applied magnetic field, \textit{n} = number of moles), as shown in Fig. \ref{Fig2}(a), significant differences between the different synthetic routes emerge.
The gas-phase grown CVT and SSVG samples exhibit Néel temperatures below 50 K, which correspond well to those of other CVT-grown samples reported in recent articles.\cite{Baenitz2021,Ag_han2022,Ag_guimaraes2024,Ag_kim2024_O}
However, the maximum of $\chi_{mol}$ for flux-grown \ce{AgCrSe2} crystals lies with 58\,K significantly higher and is comparable to that of polycrystalline samples obtained through solid-state synthesis.\cite{Ag_gautam2002} 

When analyzing the maxima of $\frac{d\chi_{mol}}{dT}$, two peaks are observed for \textbf{H} $\parallel$ \textbf{ab} (\textit{T}' = 45 K and 50 K) , while in the \textbf{H} $\parallel$ \textbf{c} orientation, only one maximum corresponding to the higher-temperature peak is observed (Fig. \ref{Fig2}(b)). This is in direct, qualitative agreement with CVT-grown crystals\cite{Baenitz2021}, with again lower temperature values reported for the crystals grown from the gas phase (\textit{T}' = 32 K and 38 K) as primary distinction.

The difference between crystals grown from various synthesis routes can also be traced from the saturation fields when considering the field-dependent magnetic moment displayed in Fig. \ref{Fig2}(c). After a linear increase (\textbf{H} $\parallel$ \textbf{c}) or one interrupted by signatures of a spin-flop transition at lower fields (\textbf{H} $\parallel$ \textbf{ab}), the saturation of magnetic moments is reached at values of ca 2.8 $\mu_B$/Cr atom. This saturation moment is in close agreement with the theoretically expected value of 3 $\mu_B$/Cr considering the spin-only formula for $d^3$-Cr\textsuperscript{III}.
The saturation fields were determined from the 50\% point of the first derivative $\frac{dm}{dH}$ during its decline to 0 after saturation at high fields (Fig. \ref{Fig2}(d)). Both flux- and gas-phase grown samples exhibit some direction-dependent differences in the saturation fields, with our \ce{AgCrSe2} crystals grown from self-flux exhibiting higher saturation fields (\textit{H}\textsubscript{sat} = 24.9 T (\textbf{H} $\parallel$ \textbf{ab}) and 26.7 T (\textbf{H} $\parallel$ \textbf{c}) than what was reported earlier for CVT-based samples (\textit{H}\textsubscript{sat} = 23.7 T (\textbf{H} $\parallel$ \textbf{ab}) and 25.1 T (\textbf{H} $\parallel$ \textbf{c}).\cite{Baenitz2021} This further indicates that self-flux grown crystals recover the magnetic energy scales characteristic of stoichiometric bulk \ce{AgCrSe2}, in contrast to the systematically reduced values observed in gas-phase grown samples. The fact that the reported saturation field for polycrystalline, solid-state synthesized \ce{AgCrSe2} (($H_{sat}$ = 25.8 T)\cite{Baenitz2021} lies right in between our two direction-dependent values, underlines again that with the self-flux method, large crystals, which behave magnetically the same way as the solid-state product, can be obtained.

\subsection{Origin of the magnetic differences}

\begin{figure*}
\begin{center}
\includegraphics[width=16.5cm]{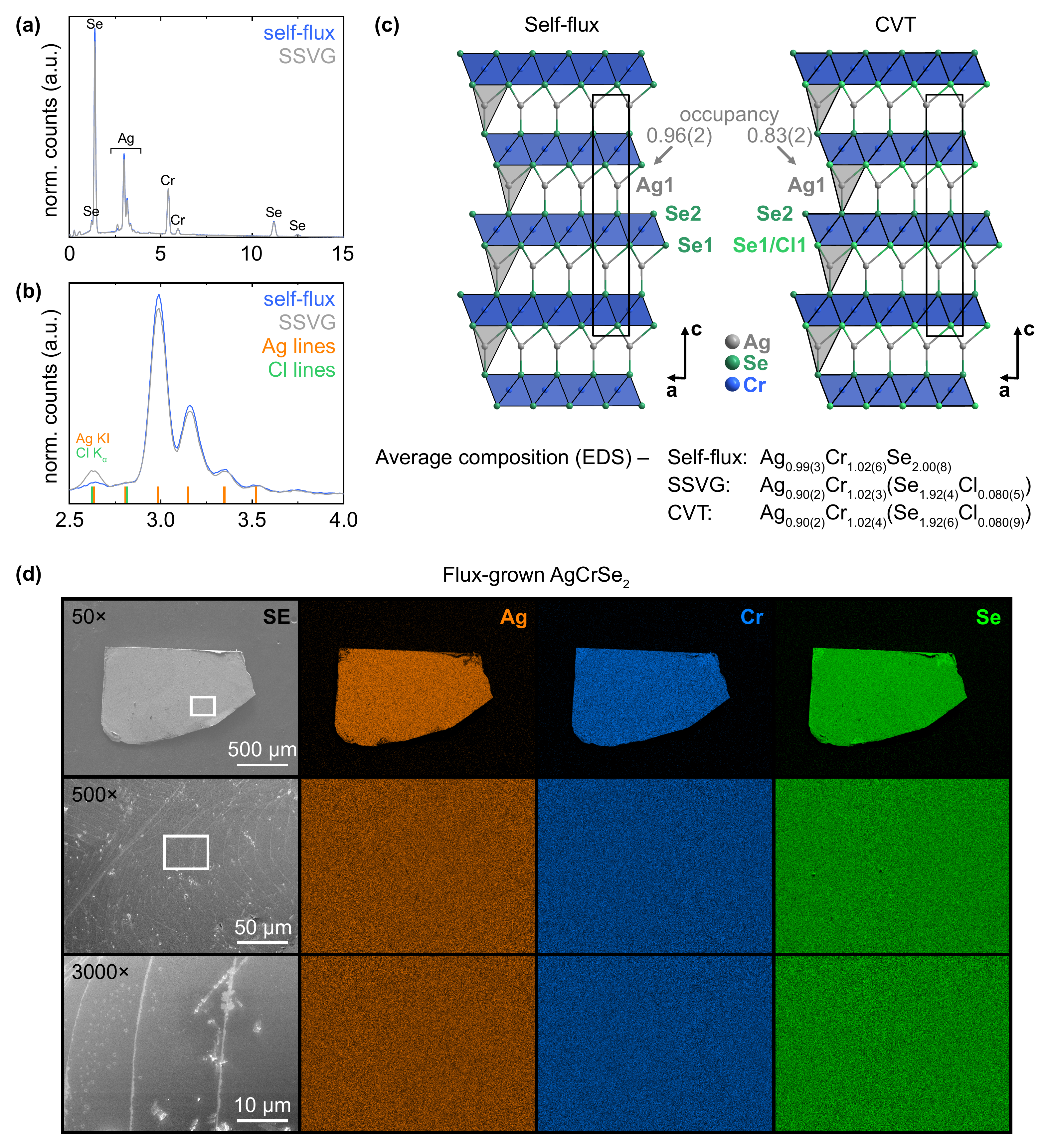}
\caption{(a) Normalized EDS-spectra of a crystal grown through self-flux and SSVG; (b) Zoom-in of the spectrum from (a) highlighting the overlap between the characteristic lines of Ag and Cl;
(c) Crystal structure of \ce{AgCrSe2} grown via the self-flux (left) and SSVG (right) technique viewed along [010]; (d) SEM analysis of a flux-grown crystal plate of \ce{AgCrSe2}. A secondary-electron (SE) picture in three different magnifications and the corresponding EDS-mappings of the constituting elements are shown; More details on the EDS data are given in the SI.}
\label{Fig3}
\end{center}
\end{figure*}

To identify the origin of the pronounced magnetic differences between gas-phase grown and self-flux grown \ce{AgCrSe2} crystals, we combine elemental analysis with single-crystal X-ray diffraction.
The observed differences in magnetic behavior cannot be rationalized from the crystal size or structure type, as both synthesis routes yield plate-like crystals with comparable dimensions and indistinguishable PXRD patterns. Elemental analysis by EDS, however, reveals a clear distinction between the compositions obtained by the different growth methods (Fig. \ref{Fig3}(a,b). Assuming a Se content of two atoms per formula unit (p.f.u.), the samples obtained through self-flux correspond to the target composition within one standard deviation: \ce{Ag_{0.99(3)}Cr_{1.02(6)}Se2} (Fig. \ref{Fig3}(a)). In contrast, crystals grown via the gas phase exhibit a significant Ag-deficiency with average compositions of \ce{Ag_{0.93(2)}Cr_{1.06(4)}Se2} (CVT) and \ce{Ag_{0.93(2)}Cr_{1.05(3)}Se2} (SSVG), when quantifying the three main elements. This is in close relation to previously reported deviations from the nominal composition (e.g. \ce{Ag_{0.92}Cr_{1.08}Se2}\cite{Baenitz2021}).

These off-stoichiometries would suggest a Cr-for-Ag substitution in the layer interspace leading to a sum formula of \ce{Ag_{1-\textit{x}}Cr_{1+\textit{x}}Se2}. Given the large portfolio of self-intercalated Cr\textit{\textsubscript{x}}Se\textit{\textsubscript{y}} or Cr\textit{\textsubscript{x}}Te\textit{\textsubscript{y}} phases\cite{blachnik1987,bensch1997}, this appears plausible, although the smaller radius of Cr\textsuperscript{III} relative to Ag\textsuperscript{I} and consequently different expected bond length is not favorable for this simple substitution.

Instead, closer inspection of the EDS-spectra of gas-phase grown samples (Fig. \ref{Fig3}(b)) reveals an enhanced intensity of the peak at 2.63 keV, which coincides with an overlap of the characteristic lines of Ag (K\textsubscript{I}, 2.633 keV) and Cl (K\textsubscript{$\alpha$}, 2.621 keV). All crystals grown by the SSVG and CVT methods, which have been analyzed by EDS, exhibited a small Cl-content of about 2\%. This appears as an inherent feature of the crystals grown from the vapor phase and could not be removed by cleaving the crystal surfaces prior to EDS analysis. Similar issues have been noticed for CVT-grown \ce{AgCrSe2} already, albeit with a slightly smaller Cl content (see SI of Ref. \citenum{Baenitz2021}).

From a crystal-chemical perspective, the most plausible mechanism for Cl incorporation into the crystal structure is by partial substitution of Se by Cl. This scenario naturally explains both the apparent selenium deficiency and the relative excess of chromium inferred from EDS when chlorine is not included in the quantification. If Cl is included in the quantification, and the molar ratios are adjusted in a way that the sum of the Se and Cl contents amounts to two atoms p.f.u., more sensible sum formulas of \ce{Ag_{0.89(2)}Cr_{1.02(4)}(Se_{1.92(6)}Cl_{0.080(9)})} (CVT) and \ce{Ag_{0.90(2)}Cr_{1.02(3)}(Se_{1.92(4)}Cl_{0.080(5)})} (SSVG) are obtained. More details on the EDS-quantification are collated in the SI. 
Moreover, we believe this partial incorporation of Cl to be the driving force for the Ag deficiency. Exchanging some of the divalent selenide anions with monovalent chloride anions reduces the formal oxidation state of the connected Cr atoms. To remain at a formal oxidation number of +III, fewer Ag counter-cations are needed in the space between the \ce{Cr(Se_{2-\textit{x}}Cl_{\textit{x}})} layers for charge-balancing. While a formal oxidation number of less than +III is in theory possible for Cr, additional \textit{d}-electrons would be associated with significantly higher energies in the octahedrally coordinated Cr atoms.  

The compositional differences among \ce{AgCrSe2} crystals grown via different methods can as well be observed in the crystal structure refinements from SXRD measurements. 
CVT-grown crystals exhibited a significantly reduced occupancy of 0.83(2) of the Ag1 site, when refining the parameter. The crystals grown from self-flux still showed a slight under-occupation of the Ag site; however, in a much less pronounced way (occupancy 0.96(2)). 
This is underlined by the fact that ignoring the under-occupation of the Ag1 site increases the $R1_{obs}$ from 0.0423 to 0.0498 for the CVT-based crystal compared to just 0.0319 to 0.0329 for the sample originating from Ag/Se self-flux. 
Even the incorporation of Cl into the crystal structure of CVT-grown \ce{AgCrSe2} can be spotted in the refinement. When freely refining the occupancy of the two Se sites, that of the one forming the base plane of the trigonal pyramidal \ce{AgSe4} coordination polyhedron exhibits light under-occupation: 0.96(3) for Se1, compared to 1.00(3) for Se2 (see Fig. \ref{Fig3}(c)). When refining this site with a partial Se/Cl occupation, 8.2(15)\% of Cl on the Se site are obtained, which is in good agreement with the observed EDS data.
For flux-grown crystals, the occupancy of the Se sites refine to full occupancy within one standard deviation (Se1 1.01(2), Se2 1.02(2)).
More details on the refinements of the SXRD data are collated in Table \ref{Crystdata}.

\begin{table}
	\begin{center}
	\caption{Crystallographic data for single-crystals of \ce{AgCrSe2} grown from self-flux and CVT}
    \label{Crystdata}
		\begin{tabular}{lll}
 \hline 
 & \textbf{self-flux} & \textbf{CVT} \\
 \hline
Chemical formula & \ce{Ag_{0.96(2)}CrSe2}  & \ce{Ag_{0.83(2)}CrSe_{1.92(4)}Cl_{0.08(4)}}\\
Mol. mass (g\; mol$^{-1}$) & 313.54 & 295.90 \\ 
Cryst. syst. & trigonal & trigonal \\
Space group & $R3m$ (160) & $R3m$ (160)\\
$a$ (\AA) & 3.68080(10) & 3.67150(10) \\ 
$c$ (\AA) & 21.2500(8) & 21.2523(7) \\ 
$V$ (\AA$^{3}$) & 249.330(16) & 248.098(16) \\ 
$Z$ & 3 & 3 \\
Calculated density (g\; cm$^{-3}$) & 6.265 & 5.941 \\
Temperature (K) & 283(2) & 293(2) \\
Diffractometer & \multicolumn{2}{l}{Rigaku Oxford Diffraction SuperNova} \\
Radiation ($\lambda$) & \multicolumn{2}{l}{Mo K$\alpha$ (0.71073 \AA)} \\
Crystal color & grey & grey \\
Crystal description & plate & plate \\
Crystal size (mm$^{3}$) & 0.16 $\times$ 0.12 $\times$ 0.03 & 0.15 $\times$ 0.09 $\times$ 0.03\\
Linear absorption coefficient (mm$^{-1}$) & 30.529 & 29.142 \\
Scan mode & \multicolumn{2}{l}{$\omega$} \\
$\theta$\textsubscript{min}--$\theta$\textsubscript{max} (°) & 2.876 -- 34.976 & 5.758 -- 34.816\\
$h$ range & --5 to 5 & --5 to 5 \\
$k$ range & --5 to 5 & --5 to 5 \\
$l$ range & --33 to 33 & --33 to 32 \\
Measured reflections & 6544 & 6580 \\
Completeness (\%) & 99.4 & 99.4 \\
Independent reflections & 340 & 336 \\
\textit{R}\textsubscript{int} & 0.0451 & 0.0637 \\
\textit{R}\textsubscript{$\sigma$} & 0.0451 & 0.0637 \\
Absorption correction & \multicolumn{2}{l} {numerical, gaussian grid} \\
Independent reflections \\ with I $\geq$ 2$\sigma$ & 336 & 330\\
$R1$ (obs / all) (\%) & 3.19 / 3.20 & 4.23 / 4.28\\
$wR2$ (obs / all) (\%) & 8.81 / 8.83 & 11.78 / 11.85\\
$GOF$ & 1.159 & 1.146 \\
Refined parameters & 16 & 16\\
Restraints & 1 & 1 \\
Maximum difference peaks ($e^-$\AA$^{-3}$) & --2.28; 2.06 & --2.63; 2.74 \\
CCDC Deposition code & 2526118 & 2526117 \\
    \hline
  \end{tabular}
	\end{center}
\end{table}

Apparently, the presence of Cl-based transport agents leads to this minor incorporation of chlorine into the crystal structure of \ce{AgCrSe2}, which we believe is the driving force for the reduced Ag-content observed in these crystals.
With the help of the self-flux method, we were able to mitigate this issue, resulting in a drastically improved stoichiometry with an average Ag-deficiency of only 1\%, which is less than one standard deviation away from the ideal 1:1:2 ratios. The elemental composition is hereby independent of the relative Ag/Cr/Se contents from the flux in a wide range. Molar Ag:Cr:Se ratios of \textit{n}:1:8 (\textit{n} = 1, 2, 3, 4, 5, 6), and \textit{n}:1:16 (\textit{n} = 6, 8) all resulted in phase-pure stoichiometric \ce{AgCrSe2}. Other than in \textit{A}/Se self-flux syntheses, which are prone to form alkali-deficient phases after quenching from high temperatures, even the Ag-poor flux compositions led to full-stoichiometric \ce{AgCrSe2}. 
As exemplified by the crystal displayed in Fig. \ref{Fig3}(d), our flux-grown crystals exhibit structural homogeneity at the large and small scope when mapping the crystal surface with SEM-EDS. While at higher magnifications, some features in the form of bent stepped terraces and small particles appear on the surface, the elemental composition remains homogeneous. 

These findings demonstrate that chlorine incorporation and the resulting Ag deficiency inherent to gas-phase growth account for the suppressed magnetic energy scales, whereas self-flux growth restores the stoichiometry required to recover the intrinsic magnetic behavior of \ce{AgCrSe2}.

\section{Conclusion}
We have established a growth method for single crystals of \ce{AgCrSe2}, which are both large and stoichiometric. For this, we have adapted self-flux synthesis paired with hot-centrifugation from mixed alkali-chalcogenide fluxes to crystal growth from an excess of liquid Ag/Se melt. Hereby, the Ag:Se ratio of the melt can be varied in a large interval (1:1--1:6) without affecting the composition of grown crystals. The resulting flux-grown crystals agree with the ideal \ce{AgCrSe2} stoichiometry within experimental uncertainty and, in contrast to gas-phase grown samples, recover the magnetic transition temperature of polycrystalline material prepared by solid-state synthesis. In addition, we demonstrate that self-selective vapor growth (SSVG) represents a viable alternative to conventional chemical vapor transport (CVT) for producing large single crystals, albeit without resolving the issue of off-stoichiometry.

While the crystal growth through CVT has enabled a multitude of advanced and direction-dependent physical characterization methods over the past decade, it has introduced significant Ag-deficiency in the crystal structure. The root cause for this lies most likely in the consequent substitution of small amounts of selenium by chlorine, originating from the \ce{CrCl3} transport agent, into the crystal structure. This off-stoichiometry can be spotted with SEM-EDS, is noticeable in SXRD refinements, where even a preferential substitution site can be identified, and significantly affects the magnetic properties in lowering both transition temperatures and saturation fields. Taken together, our findings establish self-flux growth as a route to stoichiometric \ce{AgCrSe2} single crystals with bulk-like magnetic properties. This provides a reliable platform to critically reassess whether the anomalous transport phenomena reported for \ce{AgCrSe2} single crystals are intrinsic to the stoichiometric compound or arise from the off-stoichiometry inherent to gas-phase growth methods.

\section*{Acknowledgements}

The help of Jordan Scarpetta and Tymothée Waldner in the growth of \ce{AgCrSe2} crystals is greatly appreciated. We thank Alberto Morpurgo for helpful discussions. This work was supported by the Swiss National Science Foundation under grants No. 200021-204065 and 200021-219950/1.

\section*{Supporting information}

Detailed SEM-EDS compositional analysis for all synthesis routes, magnetic susceptibility features of differently grown crystals, and SXRD refinement data including atomic coordinates and displacement parameters.

\bibliography{AgCrSe2}
\newpage

\section*{For Table of Contents only}

\textbf{Title: Resolving growth-induced off-stoichiometry in
\ce{AgCrSe2} single crystals} \\

\noindent Authors: Felix Eder, Zeno Maesen, Yurii Skourski, Enrico Giannini, Oksana Zaharko, and Fabian O. von Rohr \\

\begin{figure*}
\begin{center}
\includegraphics[width=8.89cm]{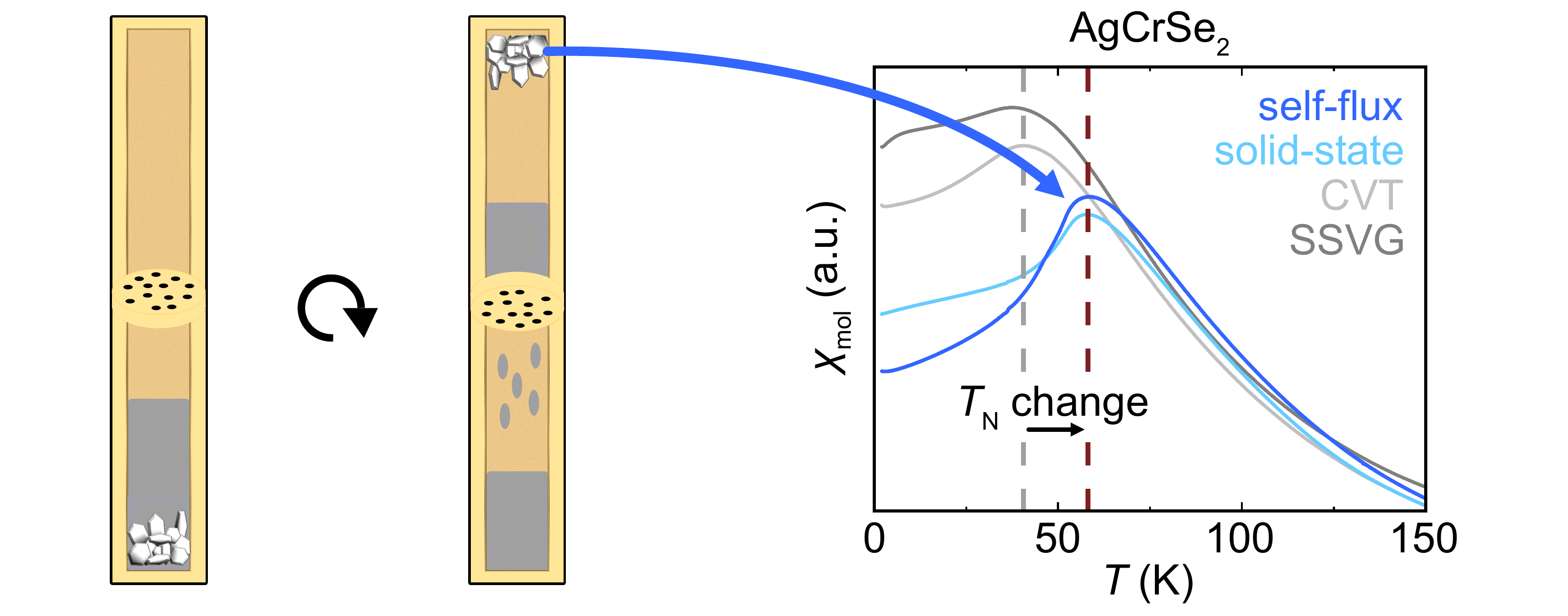}

\label{TOC}
\end{center}
\end{figure*}

 

\end{document}